# Unpriced Internal Externalities and Structural Inefficiency in Organizations

Jan van de Poll

Transparency Lab Research Institute

## Abstract

Organizations can devote substantial effort to improvement while obtaining limited aggregate benefits. This paper examines internal externalities: cross-effects of organizational mechanisms that are not fully incorporated into the decisions governing their use. We represent recurring mechanisms as primitives with signed effects across explicitly valued performance objectives. A decomposition separates effort, effect magnitude, and the extent of offsetting effects, while retaining the sign of performance contributions. We distinguish this diagnostic description from structural inefficiency, which requires a superior feasible configuration after accounting for redesign costs. Under a linear specification, we derive conditions that bound the performance attainable through effort reallocation and establish when a feasible redesign can exceed that bound. We also show why relabeling mechanisms does not improve performance and why redesign and reallocation need not be complementary. A simple extension identifies conditions for intervention sequencing without imposing a universal requirement that structural change come first. The framework connects the diagnosis of internal cross-effects to explicit comparisons of governance alternatives and provides a research design for distinguishing structural, allocative, and measurement problems. Its propositions are conditional theoretical results; the prevalence and economic importance of the proposed mechanism remain empirical questions.

## 1 Introduction

An additional approval step may prevent errors while delaying service. A common technology platform may reduce maintenance costs while restricting local adaptation. A performance target may improve the measured outcome while drawing attention away from cooperation or long-term investment. Such effects create a recurring organizational problem: a mechanism can succeed on the objective that justified its introduction while imposing consequences elsewhere. Its local success does not establish whether retaining, intensifying, or redesigning it improves organizational performance.

Economic and organizational theories already offer substantial explanations for these problems. Incentives may reward an incomplete measure of contribution, decision rights may be poorly assigned, and information may be expensive to acquire or communicate. The present paper examines a specific analytical task within that broader literature. How should an organization distinguish a weak allocation of effort within its existing arrangements from a limitation that can be overcome only by changing those arrangements?

We call a cross-effect an internal externality when it is not fully incorporated into the decision governing the mechanism that generates it. The relevant boundary is the boundary of a decision, which can lie inside a commonly owned organization. A department may evaluate a reporting requirement by the information it receives while omitting the time consumed in other departments. Common ownership makes organization-wide evaluation possible, but does not ensure that each delegated decision incorporates all consequences. Incentives, information, authority, and measurement can each explain the omission. We do not assume that cross-effects remain external when decision-makers already evaluate all consequences correctly and face the relevant organizational objective.

The distinction between an externality and a trade-off is essential. A mechanism can have positive and negative effects even when all consequences are understood and appropriately valued. Accepting slower service to satisfy a binding safety requirement may be efficient. Conversely, a mechanism can be inefficient even when all measured effects have the same sign, because a feasible alternative produces a greater benefit at lower cost. An index of opposing effects therefore cannot, by itself, establish inefficiency.

We develop a representation in which organizational primitives exert signed effects across performance objectives. A primitive is a recurring, verifiable mechanism that can be intentionally adjusted, such as an approval requirement, a coordination routine, or a technological standard. For each primitive, direction is the sum of its valued effects and magnitude is the sum of their absolute values. Internal conflict describes the proportion of magnitude offset in that aggregation. This is a descriptive property conditional on the objectives, valuations, and aggregation chosen. It is not a direct measure of hours wasted, money lost, or avoidable welfare loss.

The economic comparison enters through feasible configurations. For a given organizational structure, the best attainable performance defines an allocation benchmark. A structural improvement opportunity exists when an alternative configuration can outperform that benchmark after accounting for the resources and costs required to implement the change. These benchmarks separate what could be achieved by changing effort from what requires changing available mechanisms, their effects, or their constraints. They also prevent an unattainable configuration in which every effect is beneficial from becoming an automatic efficiency standard.

The paper makes three connected contributions. First, it provides a signed representation of internal cross-effects that can support organizational diagnosis without equating trade-offs with inefficiency. Second, it defines structural and allocative losses relative to explicit feasible benchmarks and derives sufficient conditions under which redesign outperforms any allocation within the existing structure. Third, it connects those comparisons to governance decisions, including internal valuation, unbundling, constraint, and intervention sequencing. The algebra is elementary by design. The contribution lies in the distinctions it supports and in the assumptions needed to move from a diagnostic observation to an intervention claim.

Two results delimit the argument. If a high-return alternative is already available, reallocating effort can deliver substantial gains even when the mechanisms currently used are highly conflicted. If all feasible returns are bounded, an implemented redesign can outperform reallocation when its net contribution exceeds that bound. Neither result establishes that redesign must always occur first. Redesign and reallocation can reinforce each other, substitute for each other, or be undertaken jointly. Their temporal ordering depends on implementation conditions that a static comparison alone cannot determine.

The paper is theoretical. Numerical illustrations are constructed examples, and no estimates of the prevalence of internal conflict or the relative size of structural and allocative losses are reported. Section 2 positions the analysis within organizational economics and management research. Section 3 introduces the representation and its assumptions. Section 4 develops feasible efficiency benchmarks and comparative results. Sections 5 and 6 examine governance, search, and sequencing. Section 7 presents an empirical strategy, and Section 8 concludes. Appendix A provides proofs and additional formal details.

## 2 Related literature

### Organizational economics and internal decision boundaries

The theory of the firm explains why transactions are organized within firms, how ownership and authority affect behavior, and how alternative governance arrangements perform. Coase (1937) places the costs of coordinating transactions at the center of the comparison between markets and firms. Williamson (1975) develops the role of governance in addressing contractual hazards. Grossman and Hart (1986) analyze ownership when contracts are incomplete, while Aghion and Tirole (1997) distinguish formal authority from effective control over decisions. Gibbons (2005) emphasizes differences among economic mechanisms that can support distinct theories of the firm.

These approaches do not imply that internalization eliminates all coordination problems. The present framework examines consequences that remain outside particular internal decisions and asks how incorporating those consequences changes the evaluation of mechanisms. An organizational primitive is a unit of description, not an alternative to a model of contracts or authority. The framework can be embedded in such models once the reason that some effects are omitted is specified. On its own, it does not explain the equilibrium allocation of decision rights or the persistence of an inefficient arrangement.

### Incentives and performance measurement

Jensen and Meckling (1976) analyze conflicts between principals and agents. Holmstrom and Milgrom (1991) show how the allocation of effort across tasks depends on the properties of measurement and incentive contracts. These contributions make clear that increasing the reward attached to one activity can have consequences for other activities. Ridgway (1956) similarly identifies dysfunctional consequences of performance measurement, while Kaplan and Norton (1992) argue for evaluating performance across multiple dimensions.

The present analysis treats incentive and measurement arrangements as possible sources of internal cross-effects. It does not claim that the existence of those effects is previously unrecognized. Its purpose is to separate three questions that are easily conflated: which consequences a decision incorporates, how effects are combined under an organizational objective, and whether a superior configuration is actually attainable. Better measurement may improve allocation without changing the underlying mechanisms. It changes structure only if it changes mechanisms, causal effects, or feasible choices. Recording an additional KPI does not by itself alter the production process.

### Interdependence and organizational design

Milgrom and Roberts (1990) analyze complementarities among technology, strategy, and organization. Their account makes joint choices central to understanding organizational arrangements. Levinthal (1997) examines adaptation when interactions produce rugged performance landscapes, and Siggelkow (2001) studies organizational change in the presence of fit. These approaches already provide reasons why the value of one intervention can depend on other organizational choices.

Our baseline representation is deliberately more limited. Effects are additive within a fixed configuration, and redesign changes those effects or their feasible combinations. Cross-effects across objectives are therefore distinct from interactions between primitives. A mechanism affecting several KPIs need not interact nonlinearly with another mechanism. We make that distinction explicit because neither a mixed-sign effect vector nor a large number of configurations proves that optimization is computationally difficult. The framework identifies where richer interaction models are needed rather than assuming their conclusions.

### Information attention and search

Simon (1955) provides a foundation for decision-making under bounded rationality. Ocasio (1997) examines the organization of attention, Garicano (2000) models the acquisition and communication of knowledge within hierarchies, and Sims (2003) formalizes constraints on information processing. These approaches suggest that information available somewhere in an organization may not be available, interpretable, or affordable at the point of decision.

We retain separate roles for information and search. An inaccurate estimate of a primitive's effects can produce a poor decision even when the resulting optimization problem is easy. Conversely, accurate effects may still be difficult to combine when choices are discrete, interdependent, or constrained. Neither problem is measured by the conflict index alone. This distinction also limits claims about algorithmic decision support: its contribution must be evaluated against the information, alternatives, and computational demands actually facing decision-makers.

Taken together, these literatures motivate a comparative framework rather than a general claim that existing theory neglects organizational interactions. The framework developed below provides a common representation for examining those interactions at the level of observable mechanisms. Its empirical value will depend on whether that representation improves diagnosis, predicts the effects of interventions, and adds explanatory power beyond established measures of incentives, information, and organizational complexity.

## 3 Formal framework and definitions

### Objectives valuation and primitives

Consider an organization pursuing $K$ performance objectives through $N$ recurring mechanisms. KPIs are oriented so that higher values represent more desirable outcomes. A primitive can describe, for example, the frequency of a documented review, the operation of an approval requirement, or the use of a technical standard. Its intensity is represented separately from the design of the mechanism. This distinction allows an organization to intensify an existing practice or to change how that practice operates.

Let $g$ denote the organizational configuration and let $\beta_{ik}(g)$ be the effect of one unit of effort on primitive $i$ on normalized KPI $k$. The effects refer to changes from a common baseline over a specified evaluation horizon. They are causal parameters in the model. Their empirical estimation is a separate task addressed in Section 7. We assume that they remain constant over the range of effort examined. The resulting representation is local; it should not be extrapolated beyond the range over which the approximation is credible.

We evaluate KPI changes through a stated linear organizational objective with weights $w_k > 0$. Define valued effects by

$$b_{ik}(g) = w_k \beta_{ik}(g). \tag{1}$$

The weights express the relative value assigned to normalized changes and remain fixed when comparing alternatives. Normalization alone does not establish those values. Equal weights are an explicit assumption, used in the numerical illustrations. The resulting aggregate is a measure of the specified organizational objective; it is a welfare measure only to the extent that its valuations and coverage support that interpretation. Requirements that cannot be compensated by gains elsewhere, such as minimum safety or service standards, enter the feasible set as constraints.

An improvement in this weighted objective need not be a Pareto improvement. The latter requires that no objective deteriorates and at least one improves. Similarly, an increase in the maximum weighted value after redesign does not establish that the new feasible outcome set contains the entire old set. The analysis compares performance under stated valuations and constraints. Claims of improvement across all preferences, or of an outward expansion of the full Pareto frontier, require stronger conditions. Reporting outcomes separately remains necessary when stakeholders disagree about valuations.

Let $B(g)$ collect the valued effects. Changes in $g$ can alter $B(g)$, the available mechanisms, or the feasible combinations of effort. We use a common finite index set for candidate mechanisms, with unavailable mechanisms constrained to receive zero effort. Adding a row to a questionnaire, changing a label, or adding an observation without altering choices or consequences does not constitute a structural improvement in this sense.

### Direction magnitude and conflict

Suppressing $g$ where the configuration is fixed, define direction and magnitude as

$$D_i = \sum_{k=1}^{K} b_{ik}\,, \qquad M_i = \sum_{k=1}^{K} |b_{ik}|. \tag{2}$$

Direction retains the sign of the primitive's net contribution under the stated objective. Magnitude measures the total absolute strength of its effects. A large magnitude can reflect substantial benefits, substantial harms, or both. The resultant $R_i = |D_i|$ measures absolute net strength and does not distinguish improvement from deterioration.

For $M_i > 0$, define internal conflict as

$$C_i = 1 - \frac{|D_i|}{M_i}. \tag{3}$$

We set $C_i = 0$ by convention when $M_i = 0$ and identify such primitives as inactive in the assessed objective space. The convention avoids division by zero and carries no substantive claim about the desirability of an inactive mechanism. For nonzero magnitude, the triangle inequality implies $0 \le C_i \le 1$. Conflict is zero when all nonzero effects have the same sign, including when every effect is negative.

The index has a useful alternative expression. Let $P_i$ be the sum of positive valued effects and $N_i$ the absolute sum of negative valued effects. Then

$$D_i = P_i - N_i, \quad M_i = P_i + N_i, \quad M_i C_i = 2\min(P_i, N_i). \tag{4}$$

Conflict therefore records the extent of opposition within this aggregation. It does not measure a separate causal force or a physical quantity of wasted effort. In particular, $M_i C_i$ is not generally the cost of the negative effects: when $D_i \ge 0$, it equals twice $N_i$. Treating that quantity as an externality charge would require an additional justification.

The index depends on what is measured and how it is valued. Changing a KPI weight can change both direction and conflict. Duplicating a correlated KPI can also distort the aggregate unless the valuation is adjusted. Across organizations or periods, comparison therefore requires a consistent objective system or an explicit account of differences. Reporting $D_i$, $M_i$, and the underlying effects alongside $C_i$ is more informative than reporting conflict alone.

### Effort and signed performance

Let $E_i \ge 0$ denote the effort applied to primitive $i$ in normalized resource units. Effort may represent attention, implementation intensity, or an operating input whose units are made comparable for the allocation problem. Under the local linear specification, signed performance is

$$Y_i = E_i D_i, \qquad Y(E, g) = \sum_{i=1}^{N} E_i \, D_i(g). \tag{5}$$

**Proposition 1 Performance decomposition.** For each primitive, the absolute contribution satisfies

$$|Y_i| = E_i M_i (1 - C_i). \tag{6}$$

Equation (6) follows directly from the definitions. The substantive modeling assumptions are the fixed coefficients and linear scaling in equation (5). With structure fixed, the marginal performance contribution of effort is $D_i$. Returns are constant over the modeled range. Diminishing returns, saturation, and a decline in effects at high intensity require a richer specification. A ceiling on aggregate performance additionally requires a bounded feasible set of effort.

Signed contributions must be retained when aggregating. In general, the absolute value of total performance is smaller than the sum of absolute primitive contributions. Opposition can arise within a primitive across objectives and between primitives affecting the same objective. These are different aggregation questions. Computing a row-level conflict index does not capture every form of cancellation in the full matrix, and summing row resultants can conceal negative contributions elsewhere.

For a KPI considered separately, the realized contribution of primitive $i$ is $E_i \beta_{ik}$. Any KPI-level opposition index should use these intensity-adjusted effects, with a consistent scale. The equivalence of its algebraic form to equation (3) does not establish that it measures the same organizational phenomenon. The unit of analysis must be specified before comparisons are interpreted.

### Illustrative mechanisms

Table 1 uses two normalized objectives, equal weights, and one unit of effort. The values are constructed to clarify the definitions; they are not estimates from an organization.

**Table 1 Signed effects and descriptive conflict**

| Mechanism | Effect 1 | Effect 2 | Direction | Magnitude | Conflict |
|---|---|---|---|---|---|
| A | 3.00 | -2.00 | 1.00 | 5.00 | 0.800 |
| B | 1.50 | 1.00 | 2.50 | 2.50 | 0.000 |
| H | -2.00 | -1.00 | -3.00 | 3.00 | 0.000 |
| L | 0.05 | 0.05 | 0.10 | 0.10 | 0.000 |
| A after feasible redesign | 3.00 | -0.20 | 2.80 | 3.20 | 0.125 |

Mechanism A is influential and has opposing effects, but still makes a positive net contribution. B contributes more to the stated objective despite having lower magnitude. H has zero conflict and is harmful on both objectives. L has aligned effects but a small contribution. The redesigned A would improve the net return if that configuration were feasible, but its desirability still depends on the costs and constraints considered below.

Reducing conflict is not a sufficient objective for redesign. For example, changing the effect vector from $(-4,1)$ to $(-2,1)$ improves direction from $-3$ to $-1$ while increasing conflict from $0.4$ to $2/3$. Similarly, adding another objective can increase or decrease the index depending on its sign and magnitude. These properties make conflict a diagnostic descriptor rather than a monotonic measure of organizational welfare.

## 4 Structural and allocative inefficiency

### Feasible benchmarks

Let $\mathcal{E}(g)$ be the nonempty, compact set of feasible effort allocations under configuration $g$. It incorporates operating resource limits, availability of mechanisms, and binding organizational requirements. Let $\mathcal{G}$ be a finite set of configurations admitted for comparison, including the current configuration $g_0$. Finiteness makes the benchmark transparent; the results extend to suitable compact design sets with the continuity conditions needed for an optimum to exist.

All alternatives are evaluated over the same horizon and under a stated common resource envelope. If redesign consumes resources that would otherwise support operations, the relevant feasible set must reflect that reduction. Let $\kappa(g, g_0) \geq 0$ be the direct and transitional cost of redesign in the units of the organizational objective, with $\kappa(g_0, g_0) = 0$. Costs already included as KPI effects must not be subtracted again. Resource feasibility and value accounting must be specified consistently, including the financing of redesign and any loss of performance during implementation.

Define the allocation benchmark and the best net design benchmark as

$$V(g) = \max_{E \in \mathcal{E}(g)} Y(E, g), \tag{7}$$

$$V^* = \max_{g \in \mathcal{G}} \{V(g) - \kappa(g, g_0)\}. \tag{8}$$

For an actual feasible allocation $E^0$ under $g_0$, define allocative and structural losses relative to these benchmarks by

$$L_A = V(g_0) - Y(E^0, g_0), \qquad L_S = V^* - V(g_0). \tag{9}$$

Both are nonnegative because actual effort is feasible and retaining the current configuration is an admissible option. Their sum equals the gap between observed performance and the best net benchmark. Structural loss is positive only if a feasible costed configuration can outperform the best allocation under the current arrangement. It can be zero even when conflict is high, and positive when conflict is low.

This decomposition first optimizes allocation within the current structure and then compares optimized structures. It is an accounting order, not a prescription about implementation time. If structure and allocation interact, another decomposition order can attribute the joint gain differently. The reported shares therefore depend on the benchmark and attribution convention. They are not invariant properties of the organization, and cannot be inferred from the conflict index alone.

The set $\mathcal{G}$ is consequential. A narrow set can understate available structural opportunities; an unrealistic set can overstate them. Legal requirements, technical limitations, stakeholder acceptance, and organizational change capacity may exclude an apparently attractive design. A claim that structural losses dominate allocative losses therefore requires evidence about both alternative allocations and alternative configurations. Comparing current performance with a hypothetical configuration in which every negative effect has simply disappeared does not satisfy that requirement.

### What reallocation can accomplish

**Proposition 2 Allocation with freely transferable effort.** If the only allocation constraints are nonnegative effort and $\sum_i E_i \leq \overline{E}$, then

$$V(g) = \overline{E} \max\{0, \max_i D_i(g)\}. \tag{10}$$

The organization allocates effort to a primitive with the highest positive direction, or leaves effort unused when all directions are nonpositive. This result is restricted to freely transferable effort. Minimum activity requirements, diminishing returns, joint prerequisites, or capacity limits can change the solution. Within the stated linear problem, however, examining a large number of possible allocations is unnecessary: the solution follows from comparing the coefficients.

This proposition shows why high conflict among currently emphasized mechanisms cannot establish that optimization has little value. Using Table 1, let the effort budget be ten units and actual effort be eight units on A and two on B. Actual performance is thirteen, while allocating all ten units to B produces twenty-five. The allocative loss is twelve. Reallocation provides a large improvement without changing a single effect coefficient, even though the mechanism receiving most initial effort has high conflict.

Suppose the redesign of A shown in Table 1 is feasible, retains the same operating budget, and has a cost of two objective units over the evaluation horizon. The optimized redesigned configuration produces twenty-eight before that cost and twenty-six after it. If the feasible design set consists of the current configuration and this redesign, structural loss is one. The total gap is thirteen: twelve attributable to allocation under the stated convention and one to structure. A salient conflict problem therefore coexists with a larger allocative opportunity in this example.

**When redesign exceeds the reallocation bound**

**Proposition 3 A sufficient condition for redesign superiority.** Suppose all allocations under $g_0$ satisfy $\sum_i E_i \leq \overline{E}$. Suppose also that every available primitive with nonzero magnitude satisfies $M_i \leq \overline{M}$ and $C_i \geq \underline{C}$. Then

$$V(g_0) \leq \overline{EM}(1 - \underline{C}). \tag{11}$$

If a feasible redesign $g'$ admits an allocation $E'$ for which

$$Y(E', g') - \kappa(g', g_0) > \overline{EM}(1 - \underline{C}), \tag{12}$$

its net performance exceeds that of every allocation feasible under $g_0$. The bound follows from $D_i \leq |D_i| = M_i(1 - C_i) \leq \overline{M}(1 - \underline{C})$ and the effort constraint. Additional operating constraints can reduce the attainable baseline value, so the bound remains valid when they are imposed.

The conditions control both conflict and magnitude and apply to available alternatives, including primitives receiving no effort initially. They do not follow merely from observing high conflict in the current allocation. The bound is sufficient and can be loose; failure to exceed it does not establish that redesign is unattractive. When $V(g_0)$ can be computed directly, comparing net redesigned performance with that value gives the more precise test.

For illustration, consider a different organization with only two mechanisms: A has effects $(3, -2)$ and B has effects $(1.4, -0.9)$. With ten effort units, the best current performance is ten. Allocating eight units to A and two to B produces nine, so allocative loss is one. Both mechanisms have magnitude at most five and conflict at least $0.75$, giving an upper bound of 12.5. A feasible redesign of A to $(3, -0.2)$ produces twenty-eight at full effort and twenty-six after a redesign cost of two. It exceeds the bound. With these two configurations as the design set, structural loss is sixteen and exceeds allocative loss.

The two illustrations establish possibilities, not empirical frequencies. Their different conclusions arise from different available alternatives and redesign opportunities. Whether a real organization resembles either case must be determined from its feasible choices, effect estimates, and implementation costs. Structural dominance is a comparative result to establish rather than a consequence of describing an organization as complex or highly controlled.

## 5 Governance and internalization

### Local evaluation and omitted consequences

The conflict representation describes effect patterns. To connect it to internal externalities, one must additionally specify which effects a decision incorporates. Let $\mathcal{K}_i$ be the set of objectives considered in the local evaluation of primitive $i$. Its local contribution and omitted contribution are

$$\ell_i = \sum_{k \in \mathcal{K}_i} b_{ik}\,, \qquad s_i = \sum_{k \notin \mathcal{K}_i} b_{ik}\,, \qquad D_i = \ell_i + s_i. \tag{13}$$

An omitted negative contribution makes local evaluation too favorable relative to the stated organizational objective. In Table 1, if the owner of A considers only its first effect, its evaluated contribution is three rather than one. The negative effect of two is external to that decision. Both the identity of the affected parties and the reason for the omission matter. The omission may arise from a narrow remit, incomplete measurement, delayed consequences, or costly communication; this representation does not select among those explanations.

Correcting local evaluation from $\ell_i$ to $D_i$ can change effort allocation while leaving structure unchanged. Such an improvement belongs to the allocative part of the comparison. Changes in internal evaluation can also induce subsequent redesign, but that effect requires a model of how design choices respond. Describing every benefit from internalization as structural would blur the distinction established in Section 4.

A shadow charge of $-s_i$ per unit of effort would correct the omitted valuation in this simple representation; it becomes a credit when omitted effects are beneficial. Implementing such a correction requires credible attribution, agreed valuations, and treatment of shared resource constraints. Information may be privately held and reporting strategically chosen. We therefore interpret the expression as an evaluative benchmark, not as a proof that a transfer scheme decentralizes the organizational optimum. The conflict index alone supplies neither the appropriate charge nor its allocation across affected parties.

### Unbundling and genuine changes in choice

Modularity can create valuable options for changing parts of a system, but those options depend on the design of dependencies and interfaces (Baldwin and Clark, 2000). In the present framework, the distinction between a reporting decomposition and an operational decomposition is exact.

**Proposition 4 Relabeling invariance.** Suppose a primitive with effect vector $b_i$ is represented by two components satisfying $b_i = b_i^a + b_i^b$. If the components must retain the same intensity as the original primitive, and all costs and feasible outcome combinations are unchanged, the representation leaves attainable performance unchanged. Any change in component-level conflict indices then reflects a change in description rather than an efficiency gain.

For example, the vector $(2, -1, -1)$ has zero direction and maximal conflict. It can be written as $(2,0,0) + (0, -1, -1)$. Each component has zero conflict, but applying both at the original intensity still yields the same outcomes and zero aggregate direction. A claim that this unbundling has eliminated inefficiency would confuse a change in the index with a change in organizational capability.

Unbundling becomes an intervention when it changes what can be done. It may permit independent intensities, remove a technological dependence, or redesign an interface so that a beneficial component no longer requires a harmful side effect. If every original outcome remains feasible and new options are added, gross attainable performance cannot fall. A strict net improvement additionally requires a newly feasible choice whose advantage exceeds the costs of creating and operating the separation. These costs and constraints must appear in the benchmark rather than being omitted by the decomposition.

### Selecting governance responses

The representation supports five candidate responses, but their selection depends on net value and feasibility. Scaling is appropriate when additional feasible effort produces a worthwhile positive contribution relative to alternative uses. A low conflict index alone does not establish this: mechanism L in Table 1 has little effect, and H is harmful. Moreover, a constant-return approximation will eventually require reassessment when capacity or saturation becomes relevant.

A guardrail changes feasible behavior, for example by imposing a minimum quality threshold or a limit on operational risk. Its value depends on the consequences it prevents, the flexibility it removes, and its administrative cost. Constraints can protect objectives that should not be traded away through a weighted sum. Adding a guardrail should therefore be evaluated as a change in the feasible set rather than presumed to improve the conflict index.

Unbundling is relevant when a real dependence can be relaxed at an acceptable cost. Redesign or removal is relevant when an alternative arrangement improves net value. A negative direction can identify a candidate for investigation, but removal may violate a binding requirement or omit benefits outside the measured objectives. Finally, deprioritizing a low-contribution mechanism can conserve scarce attention, provided that doing so respects its dependencies and obligations. The small direct effects recorded for one mechanism do not rule out a role as a prerequisite for other activities in a richer model.

Governance attention should consequently follow the expected value of a feasible intervention, including the value of learning more about uncertain effects. High magnitude and opposing signs can justify investigation because consequential effects are at stake. They do not establish that this mechanism must be repaired before another one. The decision requires comparing improvement opportunities, implementation burdens, and uncertainty rather than ranking mechanisms solely by a descriptive score.

## 6 Effort allocation and sequencing

### Search under a fixed structure

The baseline allocation problem separates a resource choice from changes in available mechanisms. With a common effort budget and constant coefficients, Proposition 2 has a direct solution. More elaborate linear constraints can require linear programming, but a large number of possible allocations alone does not establish intractability. Discrete choices, uncertain coefficients, interactions, and nonlinear response functions are possible extensions; each changes the problem and requires its own analysis.

Managers can nevertheless select weak allocations within a mathematically simple environment if coefficients are misestimated, objectives are incompletely represented, or relevant alternatives are absent from the decision process. Those explanations differ from a computational inability to solve the stated model. Organizational information and attention theories are therefore relevant even when the analyst's optimization problem is easy (Simon, 1955; Ocasio, 1997; Garicano, 2000; Sims, 2003).

Algorithmic support can estimate consequences, make alternatives visible, or search a difficult choice set. These contributions should be distinguished empirically. More recorded data need not provide more decision-relevant information, and adding an aligned or redundant KPI need not increase difficulty. The present model does not establish that managers require artificial intelligence to make efficient decisions. It provides quantities and counterfactuals with which a more specific claim about decision support could be investigated.

### Complementarity between redesign and reallocation

Let $E^0$ and $E^1$ be two allocations feasible under both $g_0$ and $g_1$. Write $D_i^0 = D_i(g_0)$ and $D_i^1 = D_i(g_1)$. Consider the change in the benefit of the specified reallocation after redesign.

**Proposition 5 Conditional complementarity.** Under the additive model, the interaction is

$$I = \{Y(E^1, g_1) - Y(E^0, g_1)\} - \{Y(E^1, g_0) - Y(E^0, g_0)\} \quad (14)$$

$$I = \sum_i (E_i^1 - E_i^0)\,(D_i^1 - D_i^0). \quad (15)$$

It is positive when redesign increases the benefit of that reallocation, zero when the gains are additive, and negative when redesign reduces its benefit. If redesign costs are identical under the two allocations, they cancel from this comparison. Allocation-dependent implementation costs introduce an additional interaction term and must be included explicitly.

For example, with one unit of effort and baseline directions $(1,2)$, moving effort from the first to the second primitive improves performance by one. Redesign to $(1,3)$ increases the gain from that move to two. Redesign to $(3,2)$ instead makes the same move harmful, even though the best attainable performance rises from two to three. If effort initially rests on the first primitive in this latter example, the gain from optimal reallocation falls from one to zero after redesign. Better structure therefore does not generally make optimization more valuable.

Equation (15) concerns specified choices and holds where all four combinations are feasible. When a redesign changes feasibility, a factorial comparison must distinguish an infeasible treatment combination from an unfavorable one. At the level of optimal choices, the maximizers may also change. Complementarity is an empirical and theoretical question about the relevant alternatives, consistent with the broader literature on organizational fit; it is not an automatic consequence of lower conflict.

### A limited sequencing result

A static comparison of optimized configurations does not determine when changes should occur. To clarify what is additionally required, consider a two-period extension. An organization plans to implement one redesign and one specified reallocation. Each is implemented at the start of a period, only one can be implemented per period, both orders are feasible, and both lead to the same final state $(E^1, g_1)$ in period two. Periods have equal length, values are undiscounted, and total implementation costs are the same in either order.

**Proposition 6 Sequencing with a common final state.** Under these conditions, implementing redesign first is weakly preferable precisely when

$$Y(E^0, g_1) \geq Y(E^1, g_0). \quad (16)$$

The difference between the two sequences is their first-period performance; the common final-period value and equal costs cancel. Thus a sufficiently valuable immediately implementable redesign should occur first, while a sufficiently valuable reallocation should occur first in the reverse case. For a one-unit effort budget, $D^0 = (1,2)$, $D^1 = (1,3)$, $E^0 = (1,0)$, and $E^1 = (0,1)$ provide a simple example in which reallocation should precede redesign despite positive complementarity.

Delays, learning, order-dependent costs, discounting, and irreversible changes can reverse the comparison and require an expanded model. Some organizations can make both changes together; others cannot implement the assumed reallocation until redesign makes it feasible. Proposition 6 is intentionally narrow. It establishes how a sequencing claim can be derived from timing assumptions rather than inferred from the static conflict index.

Structural diagnosis should therefore accompany allocation analysis. When a feasible design can outperform every allocation under the current arrangement, it warrants consideration even if immediate reallocation is also beneficial. The choice between acting now, redesigning first, or pursuing both depends on their feasible paths and net values. This conclusion preserves the practical importance of structure while avoiding a universal ordering that the baseline model cannot support.

## 7 Testable predictions and empirical strategy

### Measuring mechanisms and effects

Empirical implementation requires observations of mechanisms, their operating intensity, the outcomes they affect, and the alternatives an organization could feasibly adopt. Mechanisms should be described at a level at which an intervention has a stable meaning. An approval rule, for example, requires information about which decisions it covers, who applies it, its thresholds, and whether it is followed. Merely recording that an organization has an approval process may combine materially different treatments.

Observable practices can be recorded through administrative data, document review, direct observation, or behaviorally specified questions. Prior instrument development has examined alternatives to conventional employee polling and the validity of generated questionnaires (van de Poll, 2021; van de Poll, Yong, and van de Weijer, 2022). Such work concerns measurement instruments. It does not establish the causal effects of the practices on KPIs, the accuracy of a particular organization's coefficients, or the feasibility of recommended changes. Those questions require separate validation.

Ordered response categories should not automatically be treated as equally spaced quantities. A move from the first to the second state of a practice need not have the same effect or implementation cost as a move from the second to the third. Effects can instead be estimated as separate contrasts, with a common baseline and explicit exposure periods. If different transitions have different consequences, the local effect representation should be indexed by the relevant transition rather than assigning one coefficient to the entire practice.

Effort also requires a consistent definition. Managerial attention, staffing hours, expenditure, enforcement intensity, and ambition are not interchangeable observations. A study should identify the resource used in its allocation model and record changes in other resources. Reported ambition can be studied as an outcome or intention, but should not be treated as implemented effort without evidence. Comparisons of structural and allocative interventions should include the time consumed by diagnosis, implementation, and ongoing coordination.

### Causal identification and uncertainty

Cross-sectional correlations between practices and KPIs are insufficient to identify $\beta_{ik}$. Organizations adopt controls in response to risks, managers change routines after performance declines, and unobserved capability can influence both practices and outcomes. These processes can reverse the apparent sign of a relationship. Credible estimation therefore requires a design that addresses selection, timing, and confounding, such as randomized changes, a justified natural experiment, or a longitudinal comparison with defensible identifying assumptions.

Where possible, estimates used to recommend an intervention should be separated from the outcomes used to evaluate its success. Computing predicted performance and conflict from the same estimated coefficients and then showing that they obey equation (6) verifies arithmetic, not explanatory validity. Validation requires subsequent or otherwise held-out outcomes, documented treatment implementation, and uncertainty propagated from the estimated effects. A claim that redesign improves performance must be tested against those outcomes rather than against an index constructed to fall after the intervention.

Uncertainty is particularly important for mechanisms with effects close to zero or with uncertain signs. The conflict ratio can be unstable when magnitude is small. Studies should report coefficient uncertainty, sensitivity to valuations and KPI coverage, and the consequences of plausible model misspecification. A point estimate of maximal conflict is not sufficient evidence that an intervention is attractive. Recommendations should be assessed across plausible effects and costs, including the possibility that a proposed improvement violates an unmeasured constraint.

### Conditional hypotheses

The analytical identities are not empirical hypotheses. The assumptions supporting their application, the prevalence of omitted effects, and the consequences of interventions can be tested. We propose four conditional hypotheses and a measurement check that keep those distinctions explicit.

**Hypothesis 1 Omitted consequences and allocation.** Where local evaluations systematically omit material negative effects, supplying credible information about those effects will shift effort away from the affected mechanisms when feasible substitutes exist. The empirical test requires observing local evaluations before the intervention, measuring the omitted effects independently, and separating information changes from changes in incentives or feasible mechanisms. No systematic omission is implied simply by high conflict.

**Hypothesis 2 Returns to feasible redesign.** Where independently estimated effects and costs identify a redesigned configuration that exceeds a credible upper bound on baseline performance, implementing that configuration will improve net outcomes relative to allocation improvements confined to the original structure. This prediction jointly tests the effect estimates, feasibility assessment, implementation, and cost accounting. A failure of the predicted gain requires examining those components rather than treating the theoretical bound itself as an estimated causal relationship.

**Hypothesis 3 Heterogeneous gains from reallocation.** Within a fixed configuration, allocation interventions will produce larger gains where initially underused feasible mechanisms have higher net returns than the mechanisms receiving effort. Conflict concentrated among currently used primitives will not by itself predict weak allocation gains. This hypothesis distinguishes the available opportunity set from the current distribution of managerial attention.

**Hypothesis 4 Conditional interaction.** Redesign will increase the benefit of a specified reallocation when it improves the returns of the mechanisms toward which effort is shifted sufficiently to make equation (15) positive. The reverse pattern is possible. The interaction should be prespecified and assessed using common outcome valuations, rather than inferred from the fact that both interventions are individually beneficial.

**Measurement check Relabeling invariance.** A change in descriptive granularity that leaves actual mechanisms, intensities, constraints, and costs unchanged should leave measured outcomes unchanged even if component-level conflict indices change. A method that reports an efficiency improvement solely from this recoding fails the invariance requirement in Proposition 4. This check helps distinguish a useful diagnostic system from one that rewards changes in labels.

### A comparative intervention design

A feasible study can assign comparable organizational units to four conditions: existing practice, allocation support within the existing structure, a specified structural intervention, and the combination. The treatments must be operationally distinct. Allocation support changes which available mechanisms receive effort; structural treatment changes a mechanism, its effects, or its feasible use. If a treatment changes information, incentives, staffing, and structure simultaneously, its total effect can still be estimated, but it cannot identify each mechanism separately.

Assignment should account for spillovers. When units share infrastructure, managers, or approval systems, randomizing individuals may contaminate the comparison. Randomization at the relevant organizational level, or a design that explicitly models inter-unit exposure, is more appropriate. Attention and implementation assistance should be recorded across conditions so that a better-supported treatment is not mistaken for a superior structural design. Sample size and follow-up should be chosen for the relevant unit of assignment, outcome variability, and expected implementation delay.

The primary evaluation should use prespecified outcomes and feasibility requirements, with implementation costs included once. A weighted objective can support comparison, but results should also report the constituent KPIs and any binding thresholds. A favorable aggregate should not hide a violation of a mandatory requirement. Adoption, persistence, and adverse effects should be measured alongside the initial decision, because a technically attractive recommendation may not be implemented or sustained.

The combined treatment estimates a practical interaction only when the component interventions are comparable across conditions. A separate sequencing study is needed to compare redesign-first and reallocation-first paths. That study should record intermediate outcomes and delays, since equal final configurations can produce different cumulative benefits. The two-period result in Section 6 identifies the logic of that comparison without presuming which sequence will perform better in practice.

**Interpreting results and extending the model**

Observed null effects are consistent with several explanations: small true gains, implementation failure, imprecise measurement, an incorrect effect model, or constraints omitted from the proposed alternative. They should not automatically be attributed to hidden structural conflict. Similarly, a successful intervention does not establish that the same mechanism generalizes to other sectors or levels of organizational maturity. Replication should test whether effect estimates, constraints, and valuations transfer, and document where they do not.

The baseline model leaves important extensions open. Nonlinearities can represent saturation or crowding; interaction terms can represent dependencies between practices; dynamic models can represent learning, delayed effects, and the accumulation or removal of rules. Models of authority and political influence can explain why omitted effects persist despite common ownership. Each extension should preserve the distinction between measured opposition, feasible improvement, and the reasons a superior alternative is not selected.

For decision support, the next empirical question is whether a system improves the identification and selection of feasible interventions. Evaluation should distinguish improved information, improved causal estimation, wider search, and changes in organizational implementation. Comparisons with experienced managers, teams, simple heuristics, and conventional optimization methods are needed to identify the source of any advantage. The present framework supplies a structure for those comparisons; it does not establish a general performance multiplier for any particular technology.

## 8 Conclusion

Internal organizational mechanisms can create consequences that remain outside the decisions governing their use. Representing these consequences as signed effects across explicit objectives makes it possible to distinguish the strength of a mechanism from its direction and the extent of opposing effects. This representation can support diagnosis, but it becomes an efficiency analysis only when connected to feasible alternatives, resource constraints, and the costs of change.

The paper develops that connection through separate benchmarks for effort allocation and organizational design. It establishes a sufficient condition under which an implemented redesign exceeds every allocation feasible within the existing structure. It also establishes limits to the argument: reallocation can be highly valuable in the presence of conflict; relabeling mechanisms cannot create an efficiency gain; and redesign does not necessarily increase the value of optimization. The order of interventions requires assumptions about feasible paths and timing, illustrated by a restricted two-period result.

These distinctions support a more precise governance question. The relevant issue is which changes in effort, mechanisms, or constraints can improve the stated organizational objective at an acceptable cost. High conflict may direct attention to consequential trade-offs, while direction, available substitutes, and implementation conditions determine what can be gained. Internalizing omitted effects can improve allocation, induce redesign, or do both, depending on how decision-makers respond.

The empirical contribution remains to be established. Research must estimate causal effects, evaluate realistic alternatives, and observe implemented outcomes over suitable horizons. Such evidence can determine where structural limitations matter, where allocation remains the larger opportunity, and how information and computational support alter the comparison. The framework provides a disciplined basis for that investigation without treating descriptive cancellation, feasible inefficiency, and optimal intervention as the same claim.

# Appendix A Mathematical details

### Proof of Proposition 1

For $M_i > 0$, equation (3) implies $M_i(1 - C_i) = |D_i|$. Multiplication by $E_i \geq 0$ and equation (5) give equation (6). When $M_i = 0$, every valued effect is zero, hence $D_i = Y_i = 0$ and the identity remains valid under the stated convention. Writing $M_i = P_i + N_i$ and $D_i = P_i - N_i$ also gives $M_i C_i = P_i + N_i - |P_i - N_i| = 2\min(P_i, N_i)$.

### Proof of Proposition 2

Let $d^* = \max\{0, \max_i D_i(g)\}$. Every feasible allocation satisfies $Y(E, g) \leq d^* \sum_i E_i \leq d^* \overline{E}$. If some primitive has strictly positive direction, assigning the full budget to a maximizer attains this bound. Otherwise the zero allocation is feasible and attains zero. If all directions are nonpositive but some are zero, additional allocations may also attain the same optimum.

### Proof of Proposition 3

For a primitive with positive magnitude, the assumed bounds imply $D_i \leq |D_i| = M_i(1 - C_i) \leq \overline{M}(1 - \underline{C})$. A zero-magnitude primitive has zero direction and satisfies the same upper bound. Since effort is nonnegative, summing the inequalities weighted by $E_i$ establishes equation (11). The performance of the stated feasible redesigned allocation, net of its cost, exceeds this bound by assumption, and therefore exceeds $V(g_0)$. The proof does not require estimating the optimal redesigned allocation, only identifying one feasible allocation satisfying equation (12).

### Proof of Proposition 4

For the decomposed primitive at common intensity $E_i$, the vector of KPI contributions is proportional to $E_i(b_i^a + b_i^b) = E_i b_i$. By assumption, every original feasible choice has an equivalent decomposed choice, and conversely; costs and outcomes are preserved. The attainable objective values and their maximum are therefore identical. If the operational redesign instead enlarges the feasible set while retaining all original outcomes, its gross maximum is weakly greater. Strict net superiority still requires a gain exceeding redesign costs.

### Proof of Proposition 5

Substitution of equation (5) into equation (14) gives the sum of $E_i^1 D_i^1 - E_i^0 D_i^1 - E_i^1 D_i^0 + E_i^0 D_i^0$ across primitives. Factoring each term yields equation (15). The sign conclusions follow immediately. If the cost of redesign depends on the allocation, the interaction additionally subtracts the difference between the redesign costs under $E^1$ and $E^0$; its sign can therefore differ from the gross-performance interaction.

**Proof of Proposition 6**

Let $K$ denote the identical total implementation cost in either order. Redesign followed by reallocation yields $Y(E^0, g_1) + Y(E^1, g_1) - K$. Reallocation followed by redesign yields $Y(E^1, g_0) + Y(E^1, g_1) - K$. Subtracting cancels the common final value and cost, giving $Y(E^0, g_1) - Y(E^1, g_0)$. Its sign determines the preferred sequence under the stated assumptions.

**Benchmark and aggregation checks**

From equations (8) and (9), $L_A + L_S = V^* - Y(E^0, g_0)$. Nonnegativity follows because $E^0 \in \mathcal{E}(g_0)$ and $g_0 \in \mathcal{G}$ with zero redesign cost. The identity attributes the residual design gain after allocation has been optimized under the current configuration. It does not assert independence between design and allocation, or identify a causal effect from observational data.

The conflict index is invariant to multiplying all of a primitive's valued effects by the same positive scalar. It is generally not invariant to changing relative objective weights, changing the objective set, or regrouping mechanisms. For example, $(2, -1)$ has conflict $2/3$, whereas $(2, -1, -4)$ has conflict $4/7$. An additional opposite-sign objective can therefore reduce conflict by reversing the net direction. Any claim that adding objectives necessarily raises conflict requires assumptions beyond equation (3).

## Declaration of generative AI assistance

OpenAI ChatGPT assisted with drafting and revising the manuscript, checking algebraic examples, and preparing the document. The numerical examples are constructed illustrations and are not presented as empirical evidence.